\documentclass[12pt]{article}
\usepackage[utf8]{inputenc}
\usepackage{amsfonts,epsfig}
\usepackage[hyphens]{url}
\usepackage{hyperref}
\usepackage{breakurl}
\usepackage[authoryear]{natbib}

\usepackage{relsize}
\usepackage{fancyvrb}
\usepackage{amssymb}
\usepackage{geometry} 
\usepackage{sectsty} 
\usepackage{caption} 
\usepackage{subcaption} 
\usepackage[normalem]{ulem} 
\sectionfont{\sffamily\bfseries\upshape\large}
\subsectionfont{\sffamily\bfseries\upshape\normalsize} 
\subsubsectionfont{\sffamily\mdseries\upshape\normalsize}
\makeatletter
\renewcommand\@seccntformat[1]{\csname the#1\endcsname.\quad}

\makeatletter
\def\@maketitle{%
  \begin{center}%
  \let \footnote \thanks
    {\large \@title \par}%
    {\normalsize
      \begin{tabular}[t]{c}%
        \@author
      \end{tabular}\par}%
    {\small \@date}%
  \end{center}%
}
\makeatother

\title{\bf The Political Significance of Social Penumbras\footnote{We thank Yuling Yao for assistance in data analysis, Jared Lander for help with graphs, Sarah Cowan and Tom DiPrete for helpful comments, and the National Science Foundation, Institute of Education Sciences, and Office of Naval Research for partial support of this work. Data and code are at \url{http://www.stat.columbia.edu/\~gelman/research/data_and_code/penumbra_data_and_code.zip}}\vspace{5mm}}

\author{Andrew Gelman\footnote{Professor, Department of Statistics and Department of Political Science, Columbia University, New York. \protect\url{gelman@stat.columbia.edu }.} \and Yotam Margalit\footnote{Corresponding author. Associate Professor, Department of Political Science, Tel Aviv University. P.O. Box 39040, Tel Aviv 6997801, Israel. \protect\url{ymargalit@tau.ac.il}.}\vspace{5mm}}

\date{17 June 2019}

\begin{document}\sloppy

\maketitle

\begin{abstract}
To explain the political clout of different social groups, traditional accounts typically focus on the group's size, resources, or commonality and intensity of its members' interests. We contend that a group's ``penumbra''---the set of individuals who are personally familiar with people in that group---is another important explanatory factor that merits systematic analysis. To this end, we designed a panel study that allows us to learn about the characteristics of the penumbras of politically relevant groups  such as gay people, the unemployed or recent immigrants. Our study reveals major and systematic differences in the penumbras of various social groups, even ones of similar size. Moreover, we find evidence that entering a group's penumbra is associated with a change in attitude on related political questions. Taken together, our findings suggest that penumbras help account for variation in the political standing of different groups in society.

\vspace{5mm}

\end{abstract}

\thispagestyle{empty}

\section{Introduction}

Calls for changes to the U.S. immigration system have been a feature of American politics for several decades. In particular, activists have called for granting more visas for immigrants seeking to enter the country as well as for providing undocumented immigrants a path to citizenship. Yet, for years, strikingly little movement has been registered on these matters, either in terms of public opinion or in change of actual policy. In contrast, over the same time period, gay rights have undergone a major transformation in the country, with many states, and later the Supreme Court, recognizing same-sex marriage and with a large share of Americans expressing support for this change. The contrasting experience of these two hitherto socially discriminated groups---immigrants and gay people---raises a key question: What explains variation in the resonance and political salience of different social groups? 

Earlier research on variation in the political influence of social groups has focused either on differences in groups' resources or on features that affect their ability to overcome the collective action problem. In particular, studies have emphasized the group's size, as well as intensity and commonality of interests among group members as key dimensions in obtaining this goal and yielding influence \citep{mancur1965logic,offe1980two}. Our study introduces a different dimension that we argue is pertinent to explaining the political prominence of a social group: its {\em penumbra}, which we define as the set of people who have personal familiarity with members of the group, be it as relatives, friends, or acquaintances.\footnote{The term penumbra is used during a solar eclipse to describe the surrounding shades circling the dark shadow of the moon. Analogously, we consider the core social group to be the  equivalent of the full shadow, with its close social circles as its penumbra.} We conjecture that a systematic study of various penumbras can yield meaningful insight into the differences in the political salience of different social groups, including those of modest size. Moreover, taking into account changes in individuals' penumbra status can help explain shifts in their political attitudes. 


Unlike the concept of one's social network,  which refers to the contacts and relationships of a certain individual, penumbra refers to the circle of close contacts and acquaintances of a given \emph{group}. For example, two social groups of similarly modest size may have penumbras that vary in crucial ways: a group's penumbra can be large in size or small, it can be geographically concentrated or dispersed, and it can be composed of mostly of rich or poor people. Notably, the penumbras of social groups have yet to be studied. We offer a first systematic analysis of this concept and offer insight on its potential political significance. In particular, we provide evidence that  changes in individuals' penumbra status can help explain shifts in their political attitudes. 



The notion that interacting with a member of a certain social group could alter people's attitudes toward that group dates back at least to \citet{williams1947reduction} and the ``contact hypothesis'' of \citet{allport1954nature}, which, in its simplest form, holds that intergroup contact reduces intergroup prejudice. Since those early studies, scholars have accumulated substantial evidence that meaningful contact with members of an outgroup often brings about a change in attitudes toward the entire group, and that such effects extend to a broad range of outgroups and contact settings; see, for example, \cite{pettigrew2006meta}, \cite{pettigrew2008does}, and \cite{dovidio2003intergroup}. Those findings suggest that the penumbra of a social group could be consequential for the group's salience and clout. For example, people who personally know members of the outgroup may exhibit greater empathy to its traditions, sensitivities, or political demands, thus providing it with greater political influence. Furthermore, changes in the degree of the group members' exposure to others---for example, by having more people reveal themselves as belonging to the group---could change not only the size and characteristics of the group's penumbra, but potentially also alter attitudes on issues related to the group among the broader public. 

The study of social penumbras and their political effects is challenging in part because of the lack of comparable data about the people who are familiar with members of various groups of interest. We do not know much about the sizes of penumbras, or about their characteristics.  Moreover, there is limited theoretical clarity about the characteristics of a penumbra that would make the core group more politically salient or influential. 

And even if information on the penumbras were readily available, investigating the impact of membership in the social penumbra of the group on one's attitude toward the group is difficult because membership in the penumbra is not randomly assigned. People often know members of a social group because they choose to, or because they make certain decisions about where they want to live or spend their time \citep{pettigrew2006meta}. These choices then make a person more or less likely to meet members of the core group. This means that the correlation between familiarity with a group's members and attitudes on issues related to that group does not necessarily reflect a causal relationship. Familiarity may not necessarily be the trigger for a person's attitude.

In this paper we report findings from a novel study designed to better deal with those issues. We carried out a two-wave panel survey in which we asked a national sample of Americans a set of policy questions pertaining to a set of social groups (for example, assistance to the unemployed, same-sex marriage, and immigration restrictions). Later in the survey, we asked questions pertaining to their familiarity with members of these various social groups (for example, gay people, the unemployed, immigrants), probing them about the number of family members, close friends, and acquaintances they know within that group.  A year later, we asked the same respondents the  same set of questions, thus allowing us to track changes in membership in different penumbras and to investigate the empirical relationship between changes in penumbra status and attitudes on policies related to the relevant social group. 

Our panel design does not completely eliminate the concern that unobservable factors account for both penumbra status and policy attitudes related to the group. However, we find it more plausible that the change in penumbra status of an adult individual during a one-year period (most notably, a shift from not knowing anyone in that group to having a acquaintance in that group) is not caused by one's attitude on the group-related policy issue. This assumption allows us to identify the relationship between penumbra status and group-related attitudes, beyond what cross-sectional data would allow.

Our study has three main aims: first, to develop the concept of social penumbra and its relevance for analyzing or comparing the political standing of different social groups; second, to provide the first systematic set of empirical findings pertaining to the social penumbra of a wide range of groups in society; and, third, to demonstrate that the penumbra status of individuals can help account for their policy preferences on group-related issues. 

The analysis we present yields several findings of note. We find a wide variation in both the size and the shape of key social penumbras for groups of comparable size. For example, about 3\% of the U.S. population identifies as gay or lesbian, about half the share of people with a mortgage ``under water'' at the time of our survey. Yet the sizes of penumbras of the two groups go in the opposite direction, with 74\% of respondents saying they knew at least one gay person, and only  35\% reporting that they knew someone whose mortgage is under water.  It may be that some of this lower number comes from people's unawareness of the mortgage status of their friends and acquaintances, but that is part of the point:  if a group such as underwater mortgage holders is hidden to much of the general population, this will affect how their problem is perceived by the public.

Relatedly, we find that the penumbra of underwater mortgages is far more concentrated geographically than the penumbra of gays and lesbians. These differences suggest how specific policies such as same-sex marriage and debt forgiveness to mortgage owners, while directly affecting similarly sized groups, could resonate differently across the broader population. 

In addition, we find a strong empirical relationship between belonging to certain penumbra and group-related policy attitudes, for some groups (gay people, Muslims, NRA members) but not for others (welfare recipients, unemployed, underwater mortgages). Exploiting the panel design of the study, and the fact that respondents' penumbra status can change over time, our analysis suggests that correlation between penumbra membership and attitudes, where it exists, represents at least in part a causal relationship rather than just a self-selection process. People who come to know a member of a social group of which they had no acquaintances before are more likely to change their attitudes with respect to that group, as compared to other individuals. This change in attitudes is often modest and does not appear for all groups, but is discernable on average. 

The paper's findings contribute to the study of groups' influence in the political sphere \citep{grossman2001special, verba1995voice,schlozman2012unheavenly}. As noted, the bulk of this literature focuses on several key dimensions of groups, primarily their resources, size, and ability to overcome collective action problems. Our paper points to another dimension of import: the group's penumbra. By developing this concept, measuring its key features and showing its relevance for the way people form their policy views, this study highlights how a group's social penumbra can help account for its political standing. 

Furthermore, our analysis also suggests that social penumbra can help explain variation in magnitude and possibly speed of attitude change in the population. Consider the case of a group's penumbra suddenly growing, either because more people become part of the core group (for example, the unemployed in a recession or owners of underwater mortgages in the lending crisis) or because more of the core group's members reveal themselves as such to others (for example, coming out as gay). In such instances, changes in the size and shape of the penumbra can also bring about changes in public opinion on policies that affect the group in question. Moreover, if the change in penumbra size is significant enough, it could help explain why attitudes toward some groups can cascade quite rapidly.

Returning to the case of same-sex marriage, recall that as recently as 2004, the controversy over the topic was seen as a political winner for Republicans, prompting Karl Rove to put the issue on the ballot in various states in order to increase voter turnout among its many opponents. Only eight years later, the tables had turned, with Rove himself accusing President Obama of ``using the issue of gay marriage for political gain.'' This shift in public attitudes toward a certain social group, and in particular the rapid pace of change, begs an explanation. The concept of penumbra may offer part of such an explanation. 

Finally, our study also adds to the literature on the contact hypothesis and the coevolution of social networks and political attitudes \citep{pettigrew2006meta,lazer2010coevolution}. The problem plaguing most of the empirical tests of the theory has been the concern that cross-sectional analyses cannot tease out causality and selection. Using longitudinal data on individuals' contact with members of different social groups, we are able to track change in attitudes following a newly established contact. Our results provide support to the notion that familiarity with members of a group can affect attitudes, but clearly shows that such contact does not translate into consistent shifts for all groups. In the discussion section we assess the conditions under which such a shift is more likely.


\section{Social networks, personal contact and political attitudes}

The activity of social groups in the public realm is often aimed at affecting various political outcomes: policies, elections, or the advancement of ideas dear to members of the group. Substantial research therefore seeks to explain the conditions in which a group is likely to be more or less influential. One prominent strand of work argues that the influence of the group closely depends on the incentives facing its members. Large and dispersed groups often face the difficulty of overcoming collective action problems due to the incentive of each individual member to free ride on the efforts of others, thus benefitting from the outcome of interest without incurring the cost of taking action \citep{mancur1965logic}. Groups are able to overcome these problems and yield influence in instances whereby the benefits from action are particularly large, or when the group's network is tighter, thus providing stronger norms against free riding and better capabilities to sanction free riders and perhaps exclude them from the gains of coordinated action. 

Yet a group's social network can also affect its clout or influence in another way. When group members routinely interact with people outside the group, these interactions can facilitate greater understanding and sympathy to the group members' needs and interests. Perhaps the most widely known articulation of this notion is the intergroup contact hypothesis \citep{allport1954nature}, which holds that extended contact allows for learning about the group and the creation of affective ties that may bring about a change in views, in some cases also defeating prejudiced outlooks toward the group members. Indeed, positive contact experiences have been shown to reduce self-reported prejudice toward a range of socially disadvantaged groups, including blacks, the elderly, gay men, and the disabled \citep{works1961prejudice,caspi1984contact,yuker1987contact}. Research indicates that even unstructured contact can often reduce hostility toward group members \citep{pettigrew2006meta}. 

But contact with members of an outgroup are not necessarily positive. Thus, in some instances interaction with outgroup members could deepen apprehension or hostility. Furthermore, even when the contact is positive and associated with decreased hostility toward the members of the outgroup, the interaction need not lead to a change in views on policies pertaining to the outgroup. For example, \cite{jackman1986some} found that following positive contact, the affective reactions of whites towards blacks had changed, but no change was registered in white subjects' attitudes on policies geared toward combating racial inequality in areas such as housing, jobs, or education.

A large literature on the structure and impact of social networks has evolved in recent decades. Research has revealed the importance of interpersonal ties for a wide range of outcomes, including finding a job or raising capital for a new enterprise \citep{granovetter1995getting,sorenson2001syndication}. For some purposes, as in the case of job search, evidence suggests that having many superficial acquaintances or ``weak ties'' can be more effective than having fewer ``strong ties'' characterized by close and deep relationships \citep{granovetter1973strength}. For obtaining other distal goals, strong ties may be necessary. 

\begin{figure}
\centerline{\includegraphics[width=3.5in]{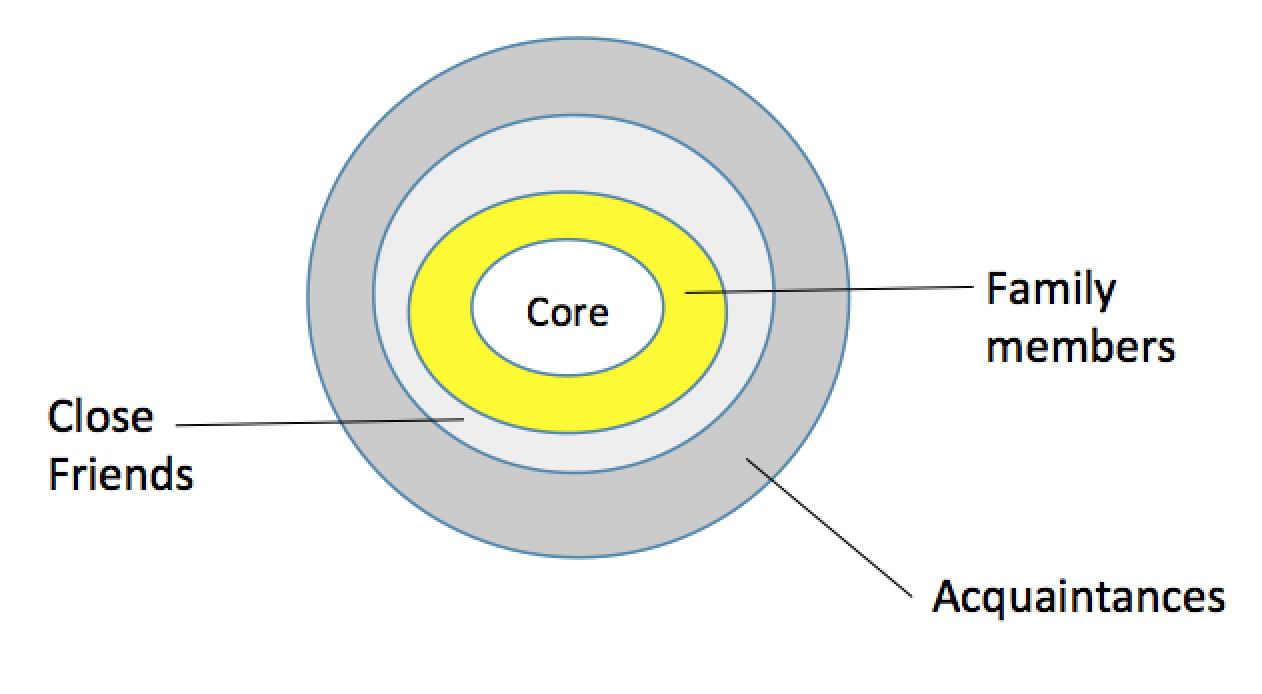}}
\caption{\em Sketch of the social penumbra of a group in the population.}
\label{fig:penumbra}
\end{figure}

In thinking about those different types of ties and contacts, and how they regulate social behavior, one can consider ``micro'' (or individual) characteristics of a social network and ``macro'' (or aggregate) characteristics \citep{jackson2008social}. Micro characteristics center on features such as whether one's friends are also friends of each other, while macro characteristics include segregation patterns and the density of links in the general social network. In the present study, we focus on the macro aspects of a network, as these relate to questions regarding the diffusion of ideas and attitudes through American society, which is characterized by a dense network of cross-cutting social ties amid political and economic polarization \citep{watts2002} .

We conjecture that an understudied factor that differentiates social groups in terms of their political significance is the structure of their social ties to others in society. We focus on the penumbra of social groups, namely the circles of meaningful relationships that members of the social group possess with their families, close friends, and acquaintances. The term penumbra is used in astronomy to describe the surrounding shades circling the dark center of the sunspot. Analogously, we consider the core social group to be the sunspot equivalent, and its close social circles as its penumbra; see Figure \ref{fig:penumbra}.  

We expect that penumbras of social groups---even ones of similar size---are likely to differ greatly in terms of their size and concentration. Furthermore, these penumbra characteristics are likely to matter politically through their association with group homophily. Homophily refers to the tendency of agents to associate with other agents who have similar characteristics to their own  in dimensions such as age, race, and profession \citep{mcpherson2001birds}. When group homophily is high, interactions are concentrated among the subpopulation. Politically, this is a significant feature of a social group, as it makes it easier for certain types of beliefs or behavior to take root among group members, but those beliefs and forms of behavior are less likely to take root or expand to the broader society than if the population were mixing evenly \citep{jackson2016economic}. 

Penumbras thus can differ with respect to the shared characteristics of the core group's members and those of their penumbra. Whereas the defining characteristic of some social groups is likely to be an important source of homophily (for example, job or sexual orientation), other characteristics are less likely to form the basis of ties with other group members (for example, the health of one's relatives, or having had an abortion). The stronger homophily is likely to be a defining feature of a core group, the stronger the ties it will have with its penumbra and the less diverse the penumbra is likely to be. 

The defining characteristic of the social group can differ in strength and meaning. For example, being a member of the group of people with green eyes is an association that is unlikely to exert a meaningful sense of belonging or shared interests, while being a member of the Muslim-American or the gay community does. This distinction raises a broader question: when is  familiarity with members of the group likely to be more meaningful in affecting people's political attitudes? Given space constraints, we shall not seek to cover the vast literature on the social identity approach and its implications for intergroup relations \citep{tajfel1971social, turner1982towards}. Suffice to say that research indicates that the impact of coming to contact with members of an outgroup differs quite significantly across cases, in part because groups are differentiated in many ways: whether the defining feature is ascriptive or whether it is linked to achievement, whether it is a majority or minority group, or the extent to which the group's defining feature is one determined by its members' choice. In what follows, we examine how this understanding translates into variation in the characteristics and political significance of different social penumbras.


\section{Data and empirical approach}

In survey research scholars are often interested in interviewing hidden, hard to reach, and marginalized populations.  Even when potential respondents are sitting out in plain sight, it can be difficult to gather a representative sample, to persuade sampled people to respond to a survey, and to get accurate responses.  What makes this even more challenging is that often researchers' interest is not just in the members of these hard-to-reach groups, but also in the way the groups relate to the larger society.


We studied penumbras using a two-wave internet panel survey designed specifically for studying this phenomenon. The survey was administered by YouGov. 3,000 respondents were interviewed in wave 1 in late August and September, 2013; of them, 2,106 were re-interviewed in wave 2 a year later. 
YouGov aims for a representative sample of American adults using quota sampling on age, sex, and other demographics.  Our wave 1 sample was unweighted, but weights are supplied for wave 2 to help deal with dropout.  In this paper we report only analyses on those respondents who completed both waves. We use survey weights when computing population proportions and averages; we do not use the weights for regression analyses that control for demographics.

\begin{table}
\centerline{\begin{small}
\begin{tabular}{lccc}
& \multicolumn{3}{c}{\% of U.S. adult population} \\
& who are in & who know & who know \\ 
Group & the group  & at least 1 & at least 2 \\\hline  
Active military & \ \ \ 0.6\% & 46\% & 29\% \\ 
Immigrated to the U.S. in past 5 years & \ \ \ 1.9\% & 18\% & 12\% \\ 
National Rifle Association member & \ \ \ 2.0\% & 41\% & 31\% \\ 
Had abortion in past 5 years & \ \ \ 2.0\% & 10\% & \ 4\% \\ 
Muslim & \ \ \ 3.4\% & 30\% & 18\% \\ 
Gay/lesbian &  \ \ \ 3.6\% & 74\% & 58\% \\ 
Lost job in past year & \ \ \ 4.2\% & 49\% & 33\% \\  
Currently unemployed & \ \ \ 4.7\% & 55\% & 38\% \\ 
Had a mortgage underwater & \ \ \ 6.6\% & 35\% & 21\% \\ 
No health insurance &  16\% & 60\% & 43\% \\ 
Spend time caring for elderly person &  17\% & 46\% & 29\% \\ 
Receive government welfare &  21\% & 49\% & 35\% \\  
Gun owner & 24\% & 77\% & 66\% \\ 
Have a serious health problem &  25\% & 74\% & 59\%
\end{tabular}
\end{small}
}
 \caption{\em Groups asked about in our social penumbra survey, listed in increasing order of their approximate sizes in the population; see Appendix \ref{groupsizes} for sources)  Group sizes were estimated using various sources on the internet.  We used our YouGov survey to estimate the size of the group's penumbra (the percentage of people who know at least one person in the group) and the percentage who know at least two in the group.}
\label{table:penumbratable}
\end{table}

We asked about penumbra membership in 14 social groups and attitude questions on 12 related policies, with some of the policy questions pertaining to more than one group. The groups include gays and lesbians, recent immigrants, National Rifle Association members, unemployed people, individuals currently taking care of an elderly family member, and others; see Table \ref{table:penumbratable}.

In selecting the set of social groups for study, we have focused on dimensions that prior literature indicates are potentially consequential for the impact that penumbra membership might exert on political preferences. These dimensions, cited earlier, include the size of the core group, whether membership of the group is voluntary or not, whether the ascriptive feature of the group is one associated with high or low social status, and whether the group is associated with the liberal or conservative camp. The advantage of exploring penumbras of groups that differ along multiple dimensions comes at a cost that we are unable to systematically assess the impact of a specific dimension that characterizes the core group---for example, whether group membership is voluntary---on the significance of the penumbra membership. This is a tradeoff we believe is acceptable given the objectives of the project:  as this is a first attempt to explore the characteristics of different social penumbra, our emphasis is instead on analyzing penumbras of groups that differ on a range of theoretically important dimensions.

Penumbra membership was constructed by asking the respondents to report the number of people from the social group who are (i) close family; (ii) close friends; (iii) other people they know. To add clarity, we defined the third category as ``people that you know their name and would stop and talk to at least for a moment if you ran into the person on the street or in a shopping mall.'' Finally, we prompted the survey respondents with eight first names and asked them to count the number of people they knew with each name. 
The addition of names helps us measure the size of each respondents' social network and also provides a check on the face validity of our estimates regarding the penumbra of the different social groups.  We asked about the names Rose, Emily, Bruce, Walter, Tina, and Kyle, chosen to represent a balance of male and female, young, middle-aged, and old; along with Jose and Maria to target the Hispanic population.

This method for eliciting information about people's close social ties allows us to generate an estimate of the absolute numbers of the different penumbras.  As we show below, the penumbras of acquaintances are larger than those of close family and close friends, which is no surprise. Yet the total numbers we calculate for the different penumbra are relatively low compared to other research on social networks.  For example, \citet{zheng2006many} estimate an average network size of about 700.  For a group such as gays that is 3\% of the population, this would correspond to an average penumbra size of 21. Yet the figures we obtain are lower; in the case of the gay penumbra, our figure is just above 5, as we will show in Figure \ref{fig:party}. This is not entirely surprising: for a number of reasons, we might expect  that survey responses would provide lower counts for some penumbras. Most obviously, it could be difficult to recall 20 people in one's social network that belong to a certain core group; for another, not all gays are ``out'' to all their acquaintances. We suspect that asking the question in the way we did motivated survey respondents to give relatively low numbers for the Other category of acquaintances, as they had just been anchored by the low numbers that they had given to the Close Family and Close Friends categories.  This does not annul the validity of our results; rather, we think of these numbers as representing the counts of near rather than distant acquaintances, and it is consistent with the general finding that averaged responses to ``How many people do you know?''\ can depend strongly on how the question is asked.

\begin{figure}
\centerline{\includegraphics[width=\textwidth]{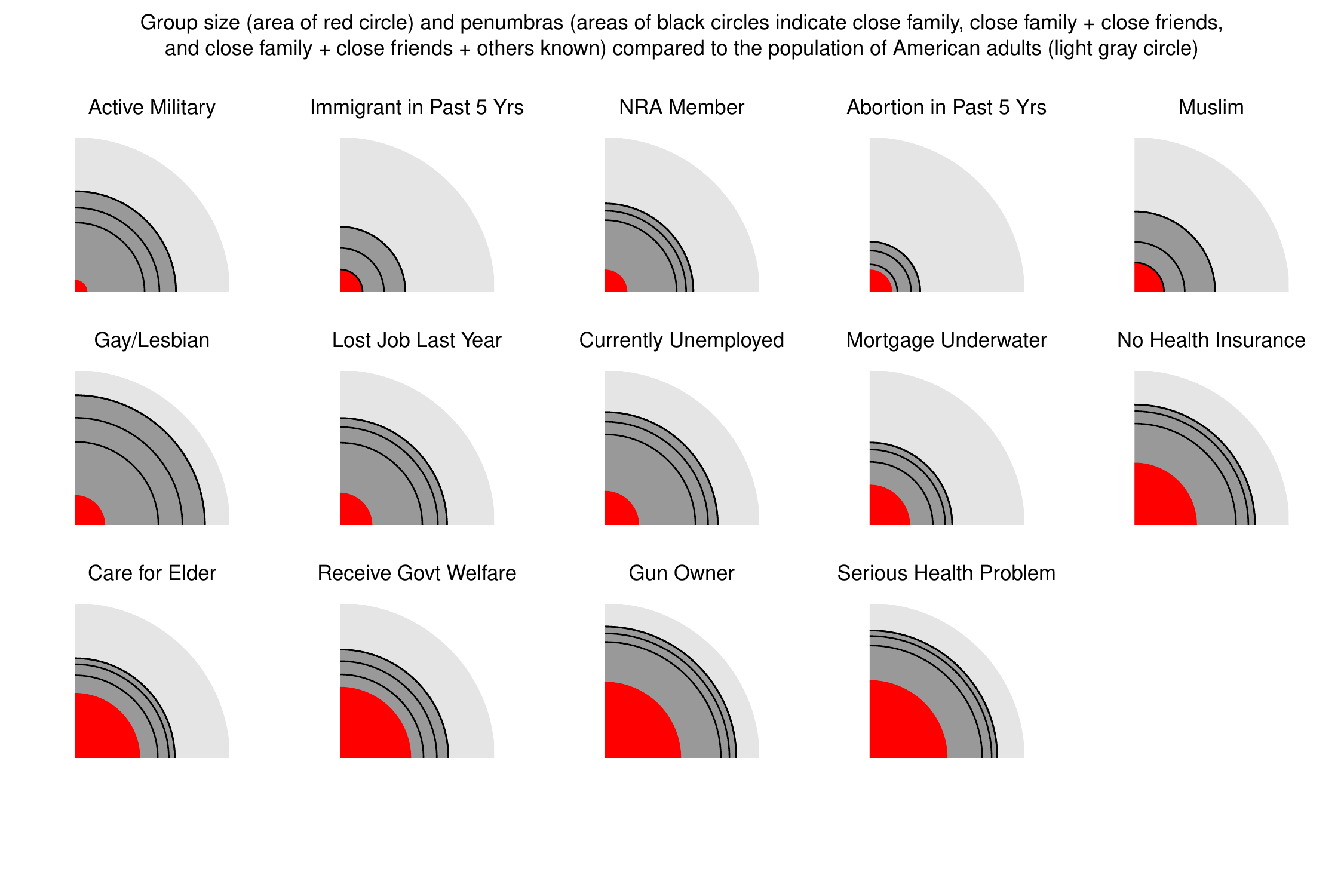}}
\vspace{-.7in}
\caption{\em Groups asked about in our survey, in increasing order of size.  For each group, the area of the red circle indicates its size, and the concentric circles show its estimated penumbra:  the percentage of survey respondents who report knowing at least one family member, close friend, or other acquaintance in the group.  Sizes and shapes of penumbras vary dramatically.}
\label{fig:bullseye}
\end{figure}

The policy questions we asked were directly relevant for at least one of the social groups of interest. For example, with respect to the immigrant penumbra respondents were asked for their views on whether more or less immigrants should be permitted to live in the United States. To assess the policy attitudes of the gay penumbra, we asked respondents for their views about gay marriage. To tap into the views associated with knowing a gun owner, we asked all respondents about their position on a nationwide ban on assault weapons. On the eldercare penumbra, we asked about tax breaks for family expenditures on care provision for elderly family members. The exact wording of all the questions appears in Appendix \ref{surveyitems}.

\section{Size and shape of social penumbras}

\subsection{Penumbra size}

Table \ref{table:penumbratable} shows the groups we asked about and our estimates of their penumbras, as measured by the percentage of respondents who knew at least one or at least two people in a group. Figure \ref{fig:bullseye} illustrates graphically, concentric circles starting with close family and then extending to include close friends and acquaintances.

The penumbra is typically much larger than the group itself. For example, less than 1\% of American adults are in the active military but nearly half of respondents know someone in the service.  This figure probably represents some combination of knowing people through one's social network as well as uncertainty about classification; for example, one might count a friend who is no longer in active service, and this would be part of the calculation of the penumbra but not in the group size.

As the table indicates, the gay/lesbian penumbra is particularly large in comparison to the size of the group itself, with nearly three-quarters of respondents reporting that they know someone among this group, which is generally estimated to comprise about 3\% of the population. This sizable penumbra may have political repercussions, as suggested by the rapid gain in acceptance of same-sex marriage in recent years.

Other groups with a similarly large penumbra include gun owners and people with serious health problems, yet note the difference in the size of the core group itself: gun owners, for example, constitute about 24\% of the population, a figure that is about eight times the number of gays and lesbians in the United States.

At the other extreme, few people report knowing someone who had an abortion in the past five years, despite there being millions of women who fall into this category.  Women who have had abortions do not always reveal this fact to their acquaintances \citep{cowan2014}. Thus, the size of a penumbra can vary greatly not only due to differences in the size of the core groups themselves, but also because of differences in the extent to which group members reveal (or can be identified) as such, which in turn can be relevant when considering public opinion and political change.

\subsection{Geographic variation of penumbras}

Another important factor that could account for variation in the size of a penumbra, controlling for the core group's size, is its spatial dispersion. Geographic segregation is a key concept in the analyses of social networks \citep{jackson2008social}. Our survey does not have data at the neighborhood level but we can get some handle on  the degree to which social groups differ in the geography of their penumbras by fitting for each group a simple hierarchical  model predicting penumbra membership across the 50 states.  We fit the model, $\mbox{Pr}(y_{ij}=1)=\mbox{logit}^{-1}(\alpha_{{\rm state}[i], j})\mbox{ for } i=1,\dots,N, \ \alpha_{s,j} \sim \mbox{N}(\mu_j,\sigma_j^2)\mbox{ for } s=1,\dots,50$, where $y_i=1$ if survey respondent $i$ is in the penumbra of group $j$, $\mbox{state}[i]$ is an index variable for the state of residence of person $i$, and we are estimating hyperparameters $\mu_j,\sigma_j$ for each group.  We fit the model in the Bayesian inference package Stan \citep{stan2019} so as to estimate the proportion of people in the penumbra, for each group and each state.  We also estimate the standard deviation of these state proportions.

\begin{figure}
\centerline{\includegraphics[width=\textwidth]{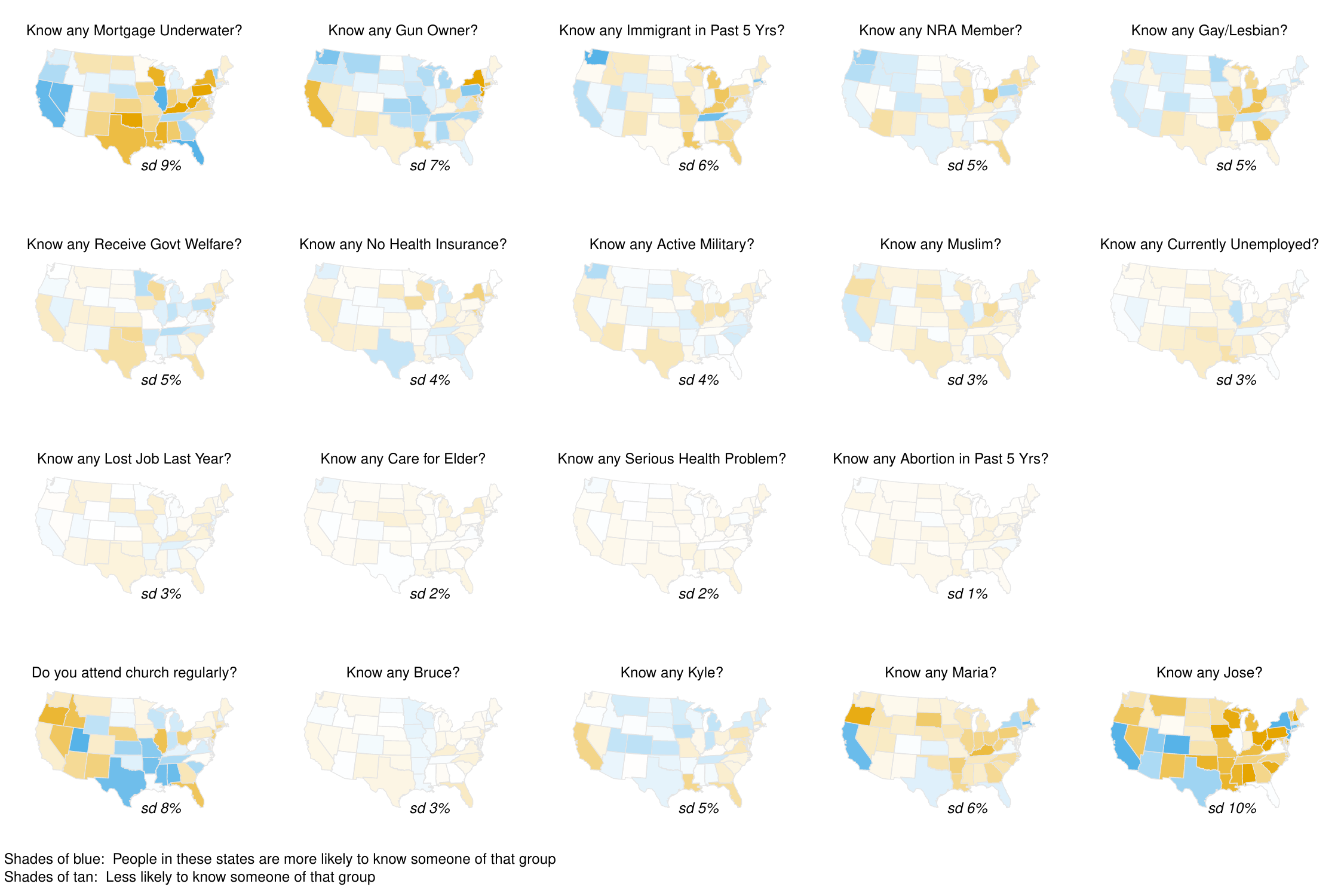}}
 \caption{\em {\em Top three rows:}  Geographic dispersion of penumbras, as estimated using a simple hierarchical model for each group.  Most of the groups' penumbras are roughly evenly distributed across the country.  {\em Bottom row:}  Geographic disperson of several survey responses.  These graphs are helpful in calibrating our understanding of the penumbra maps above.  Church attenders are more prevalent in the south and less in the west and northeast; Kyles are more common in the middle of the country, and people living in states with more Latinos are, unsurprisingly, more likely to know Marias and Joses.}
\label{fig:geography1}
\end{figure}

The top three rows of Figure \ref{fig:geography1} show, for each group, the posterior mean of the relative size of each penumbra by state, along with the posterior mean of standard deviation of the corresponding membership probabilities across states.  Before discussing the results, we emphasize that the estimates for individual states are noisy.  

The penumbra maps are shown in decreasing order of geographic dispersion.
Some groups such as underwater mortgages, gun owners, and immigrants show a fair amount of geographical variation in their penumbras. Many people know underwater mortgage owners in the high-delinquency states such as California, Nevada, and Florida, while only few know members of this group in states that did not experience the real estate bubble, such as Wisconsin, Mississippi, or Kentucky. In contrast, penumbras of other social groups such as eldercare, seriously ill and women who had abortions are essentially uniformly distributed across the country.  Again, these maps should be taken as showing general patterns, and particular state estimates can be noisy.

For comparability, we performed the same analyses and created the same maps for a group that is known to have significant geographical variation---frequent religious attenders---as well as for the penumbras of four of the first names in our study. As the bottom rows of Figure \ref{fig:geography1} show, there is appreciable cross-state variation among church attenders, while there is little geographic concentration of the penumbra of the names, with the unsurprising exceptions of Maria and Jose.

\subsection{Socioeconomic characteristics of penumbras}

\begin{table}
\centerline{\begin{small}
\begin{tabular}{lcccc}
 & Top third & College & Non-Hispanic & \\
Penumbra of \dots & of income & educated & white & Male \\\hline 
Receive government welfare & 16\% & 24\% & 69\% & 46\% \\ 
No health insurance & 19\% & 27\% & 67\% & 47\% \\ 
Currently unemployed & 20\% & 29\% & 69\% & 47\% \\ 
Lost job in past year & 21\% & 30\% & 71\% & 46\% \\  
Have a serious health problem & 22\% & 29\% & 72\% & 46\% \\
Spend time caring for elderly person & 23\% & 30\% & 71\% & 49\% \\
Had abortion in past 5 years & 24\% & 30\% & 67\% & 35\% \\   
Gun owner & 24\% & 29\% & 76\% & 48\% \\
Gay/lesbian & 25\% & 31\% & 71\% & 45\% \\  
Active military & 26\% & 31\% & 71\% & 45\% \\  
Had a mortgage underwater & 27\% & 31\% & 75\% & 46\% \\  
Muslim & 29\% & 40\% & 65\% & 53\% \\  
National Rifle Association member & 29\% & 35\% & 81\% & 55\% \\   
Immigrated to U.S. in past 5 years & 34\% & 44\% & 70\% & 51\% 
\end{tabular}
\end{small}}
\caption{\em Demographic characteristics of different penumbras, ordered by increasing percentage of upper-income people.}
\label{table:characteristics}
\end{table}

In addition to the size and (geographical) shape of the different penumbras, we also examine the variation in their socioeconomic characteristics. These include the share of high income individuals in each group's penumbra, the share of whites, of males, and the corresponding figure of college educated individuals. As Table \ref{table:characteristics}  indicates, the heterogeneity is substantial.  For example, the penumbras of welfare recipients, uninsured people, and unemployed people have the lowest share of upper-income individuals (20\% or less) and also the lowest share of college-educated respondents.  On the opposite end, the penumbras of Muslims, NRA members, and recent immigrants have a particularly high share of upper-income and well-educated people.  Such variation in the composition of groups' penumbra---in this case, the share of high income people --- may help explain why some groups have greater clout or salience in the public discussion. 

There is less variation in the race and sex breakdowns in Table \ref{table:characteristics}, with the exceptions that whites are overrepresented in the penumbras of underwater mortgages and NRA members, and women are overrepresented in the abortion penumbra, with the latter partially explainable by the fact that abortions are often kept secret \citep{cowan2014}.


We are particularly interested not just in demographic composition but also in the distribution of political attitudes within different penumbras, and we turn to that next.

\section{Penumbra membership and political attitudes}

\subsection{Correlations with party identification and issue attitudes}

\begin{figure}
\centerline{\includegraphics[width=\textwidth]{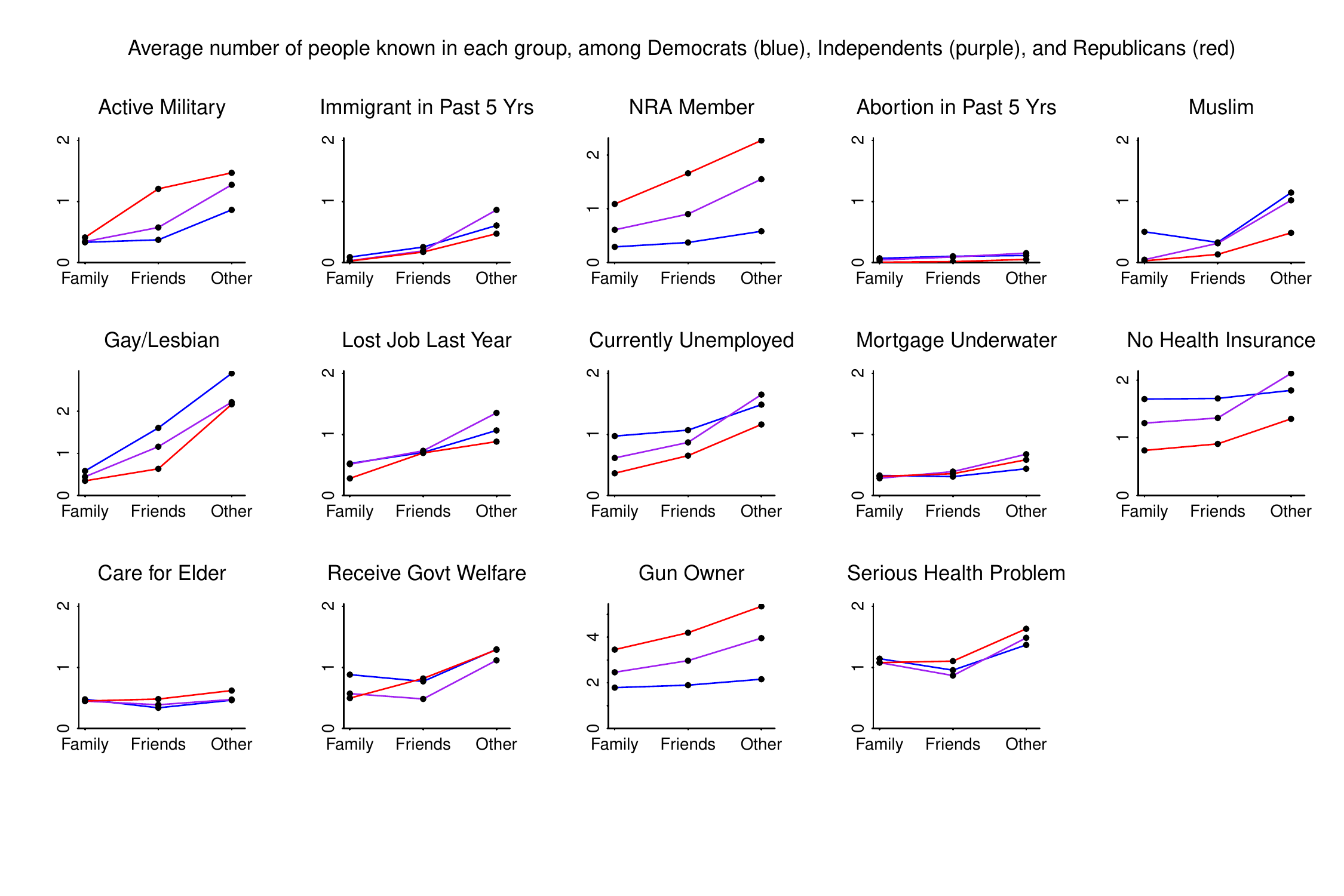}}
\vspace{-.7in}
\caption{\em Average number known in each group among respondents classified by self-reported party identification.  Differences by party are generally larger among friends and acquaintances than among close family members.}
\label{fig:party}
\end{figure}

To what extent does being part of a certain social penumbra affect one's political attitudes? In this section we assess the evidence in support of our contention that social penumbras can help account for systematic variation and change in political attitudes. We begin by exploring two sets of correlations that suggest that penumbra membership is a significant factor in shaping political affiliations. First, we analyze whether social groups differ systematically in terms of the partisan leanings of their penumbra. 

Figure \ref{fig:party} shows average penumbra sizes as a function of party identification.  Indeed, patterns are what one might expect, with Republicans knowing more NRA members and people in the military and Democrats knowing more gays and people who have no health insurance.  The larger partisan differences are among friends and acquaintances, not among close family, a finding that makes sense given that people cannot choose most of their family members.

\begin{figure}
\centerline{\includegraphics[width=\textwidth]{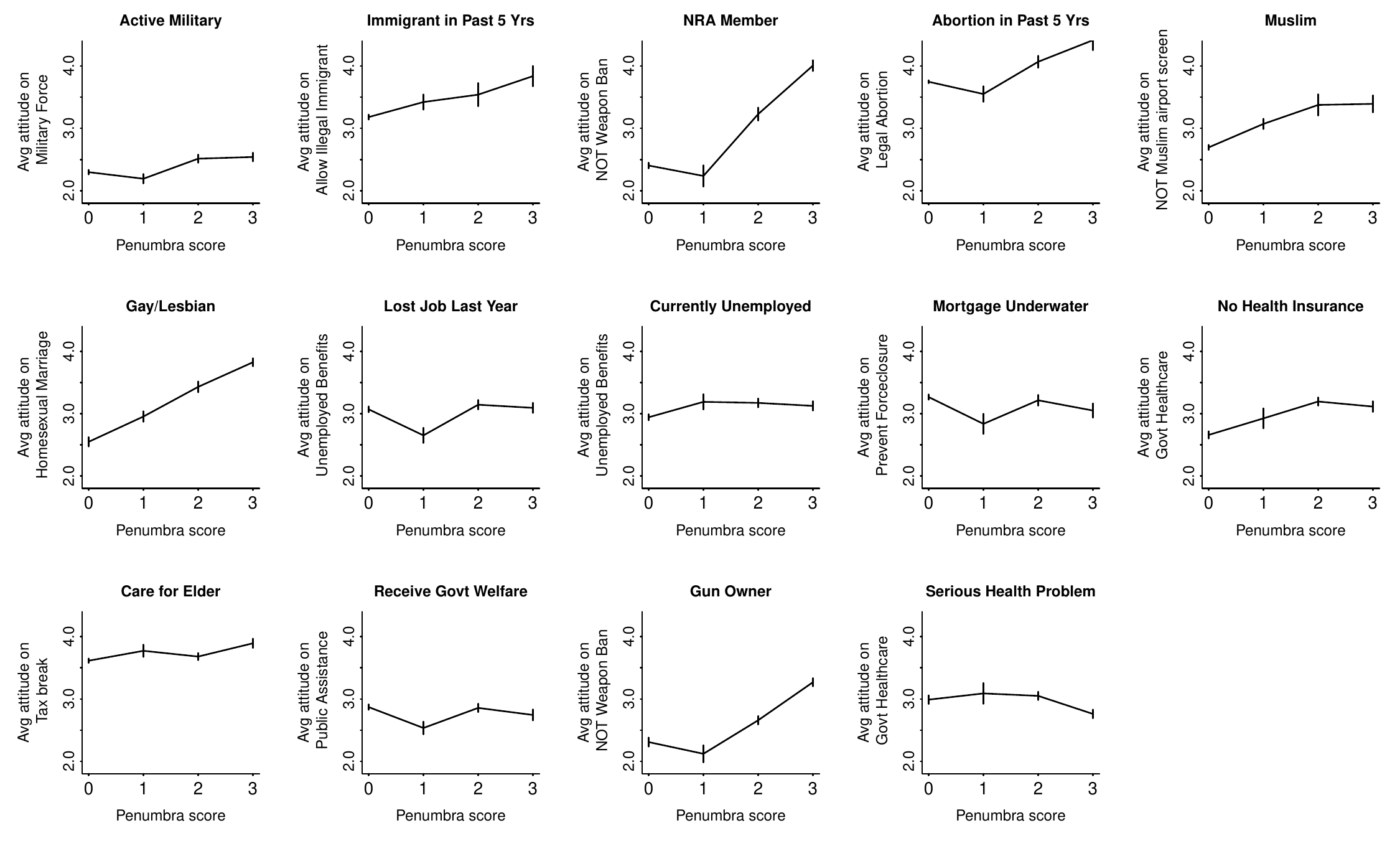}}
\caption{\em Average response on 1--5 scale for issue attitudes related to the groups being studied, as a function of the penumbra score (0 for respondents who know no people in the groups, with 2 points given for knowing a close friend or family member and 1 point for knowing at least one acquaintance.  For many but not all categories, being in the penumbra is correlated with a more positive view on the related political issue.}
\label{fig:attitudes}
\end{figure}

Next, we assess the association between penumbra membership and attitudes on policy questions directly related to the social group of interest. Figure \ref{fig:attitudes} shows the raw correlations of penumbra membership with attitudes on these policy issues.  As the figure indicates, the results vary:  The correlation between knowing NRA members and opposing a weapon ban, or between knowing gay people and supporting same-sex marriage, is strong.  For other issues the correlations are lower, and for some the correlation is zero.  Overall the correlations tend to be larger for social issues, and are much smaller (to non-existent) for attitudes on economic issues such as unemployment benefits, mortgage foreclosures, tax breaks for caregivers, and public assistance.

This pattern is consistent with a range of findings in the study of American politics and public opinion, which find that attitudes on economic issues are strongly aligned with partisanship and political ideology \citep{bartels2002beyond,margalit2013explaining}. 
Hence, it makes sense that personal contacts are less important, as compared to one's ideology, in predicting economic attitudes.  That said, we still find it surprising that there is zero correlation between knowing someone with an underwater mortgage and supporting mortgage relief, or between knowing someone with serious health problems and views on healthcare spending.  We also see almost no correlation between the active military penumbra and support for the use of military force; here the non-pattern could perhaps be explained by a mix of effects, that survey respondents in the military penumbra may feel more positive about the military in general while being reluctant to support policies that bring active duty personnel in harm's way.

\subsection{Consequences of entering the penumbra}

\begin{figure}
\centerline{\includegraphics[width=.6\textwidth]{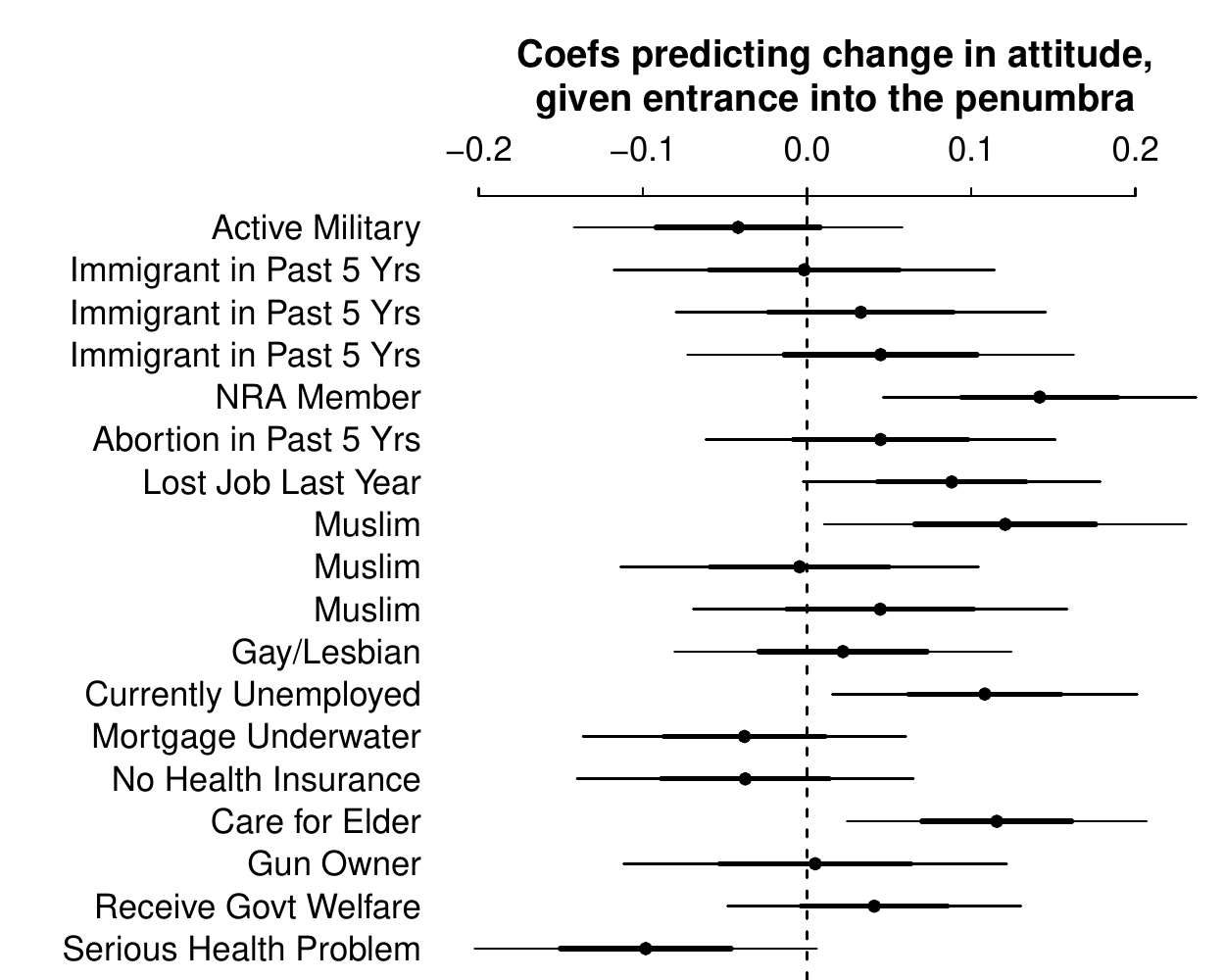}}
 \caption{\em Estimated effects with $\pm 2$ standard error bounds on particular issue attitudes, comparing respondents who entered the penumbra between the two waves of our study, to those who were not in the penumbra in either wave.  On average the coefficients are positive, and the results are consistent with an overall effect of about 0.05, but there is not enough information in the data to precisely estimate these changes for individual survey questions.  The immigrant and Muslim groups are listed three times because for these groups we looked at attitudes on three issues:  number of immigrants, illegal immigration, and airport screening for Muslims.}
\label{fig:coefs}
\end{figure}

As noted earlier, it is difficult to interpret static correlations between attitudes and penumbra membership.  We therefore seek to get more leverage on causality by exploiting the panel aspect of our study, examining \emph{changes} in attitude that occur following changes in penumbra status.  So, for each of our group penumbras and for each corresponding political issue question, we run a regression predicting change in issue attitude, given change in penumbra status and also controlling for attitude at wave 1 and several demographic and political background variables of age, sex, education, party identification, and ethnicity of respondent.
Our interest is in measuring the effect on issue attitude of entering the penumbra. Hence, we restricted the sample to respondents who were not in the penumbra in wave 1.  Entry into the penumbra is not randomly assigned, and our identifying assumption is that penumbra entry in a given period is orthogonal to the propensity to change attitude on the relevant policy question during that same time period. In many instances,  hough surely not in all, this seems to be a plausible assumption.


Figure \ref{fig:coefs} shows the results.  The scale tells us that the estimated effect for each penumbra/issue combination is small.  On average we see a positive effect (the issues have been aligned so that we expect to see a positive coefficient in each case) of about 0.05, consistent with entrance to the penumbra having a 5\% chance of shifting a respondent by 1 point on the 1--5 scale.  The wide confidence intervals indicate uncertainty about the relative magnitudes and even the signs of the effects of entering different penumbras.
The groups for which the coefficients seem most clearly positive are NRA members (and attitudes toward bans on guns), the Muslim penumbra (and opposition to airport screening), currently unemployed (and attitudes on unemployment insurance benefits), and elderly care penumbra (and support for tax breaks for caregivers).

Considering all the estimates in Figure \ref{fig:coefs} together, there is no evidence that any of the effects are negative; rather, it appears that there are small average effects of entering the penumbra, but with available data it is essentially impossible to reliably identify which effects are larger than others.  Beyond this, there can also be interactions.  For example, meeting someone who has lost his or her job in the past year might have a larger effect for those people who do not already know someone who is currently unemployed; or meeting a gay person can have a larger effect for Republicans than for Democrats.  Given the size of the effects and the degree of uncertainty in the estimates shown in Figure \ref{fig:coefs}, it is essentially impossible to estimate such interactions, but we should be aware of their potential presence.

\subsection{Robustness checks} 

We perform placebo checks to address four possible threats to identification in the panel study.

Our first concern is that the patterns in Figure \ref{fig:coefs} could simply be explained as an artifact of measurement error in penumbras.  More specifically, the issue would be as follows:  in those regressions, we are comparing people who enter the penumbra to those who stay outside.  But suppose that reports of number known in a group have some noise.  Then, even in the absence of any true effect of entering the penumbra, those people who go from reporting zero to a positive number would know more people in the group, on average, than those who remain at zero.  That is, in a hypothetical null world in which underlying penumbra membership is not changing at all but where responses are noisy, a positive reported number known in wave 2 would still be an indication of a higher probability of being in the penumbra.

To address this possibility, we re-fit our model for each group in time-reversed order, predicting change in attitude from waves 2 to 1 from change in penumbra from waves 2 to 1, just considering respondents who reported zero people known at time 2, and controlling for demographics as before.  If the result is a measurement error artifact, we should expect to see basically the same results as before, representing a residual correlation but no change or evidence of causation.  If the result is truly a change, we should not expect to see anything in the time-reversed analysis.  When we re-fit in this time-reversed order, we see no patterns beyond noise, thus suggesting that the results of Figure \ref{fig:coefs} are not explainable simply as a measurement-error artifact.

A second possibility is that changes in the attitudes could reflect nothing more than general liberal-conservative ideology, for example, that entering the gay penumbra is correlated with having more liberal attitudes more generally which would then show up as increased support for same-sex marriage, without being specific to the gay penumbra. To check this hypothesis, we perform $14\times 12$ regressions, predicting change in each of 12 attitude questions given entrance to each of 14 penumbras.  
Of the 168 regressions, 18 correspond to issue attitudes and group membership that we predicted ahead of time and 150 are cross-correlations for which we had no expectation of finding effects.  The estimated coefficients are shown in Figure \ref{figure:cross.regs}. The 150 cross-correlations show no apparent patterns beyond what might be expected by noise, a pattern we also confirm in a formal check.  For each of these groups of coefficients, we compute the sum of squares of the $t$-scores (estimate divided by its standard error) of the coefficients.  For the group of 18, this sum of squares is 35, much higher than would be expected from the $\chi^2_{18}$ distribution that would occur due to chance alone.  In contrast, for the group of 150, the sum of squares of $t$-scores is 158, completely consistent with the $\chi^2_{150}$ noise distribution.  This does not mean that there is nothing going on in these 150 regressions, merely that whatever is happening is overwhelmed by noise. This finding serves as a useful placebo check that allows us to reject the hypothesis that the positive coefficients in Figure \ref{fig:coefs} are nothing but a manifestation of a general correlation with liberal-conservative political attitudes. 

Third, we perform basic robustness checks by repeating our analysis using different specifications, considering other measures of penumbra membership (change in penumbra size or change in square root of penumbra size, instead of a binary in/out measure), excluding respondents over the age of 65 or over the age of 40 (to search for the possibility that effects could be larger among young people whose attitudes are more malleable), including an interaction of treatment effects with age, and controlling for total network size (as estimated by the sum of responses to the first names questions).  The results for those alternative analyses are broadly similar to those shown in Figure \ref{fig:coefs}:  the rankings of the different estimates change, but the average effect estimate remained positive.

Fourth, we test for a ``demand effect,'' whereby individuals who report in the second survey knowing someone from the core group (entering its penumbra) perceive the socially desirable thing to do is to report a policy attitude more favorable to the group. If this is so, the change we observe in attitudes does not reflect a true shift in attitudes but instead is an artifact of the survey design. 
 
To address this possibility, we estimate the same models as before, but replacing the outcome variable of actual observed change in position on a given policy question with respondents' \emph{self-perceived} change in position on that same question. We measure this perception based on respondents' answer to the the following question in the phase 2 survey: ``On each of these items, please try to think how your current view compares to the one you held 12 months ago.'' Respondents were then provided with options from which to describe the change in position (``I became more in favor,'' ``My opinion hasn't changed,'' or ``I became more opposed''). Due to space constraints in the survey, this question was asked for only six of the policy items. 

By using respondents' responses as the outcome, we can compare the effect of entering a penumbra on actual versus perceived change in views. When estimating the model using perceived changes, the estimated effects of penumbra entry on perceived change in policy views are all close to zero and indistinguishable from noise; see Figure \ref{fig:appx.perceived}. This supports the idea that the actual change in policy views we saw in Figure \ref{fig:coefs} arose from entering the penumbra rather than simply reflecting concerns about social desirability.

\section{Discussion}

In 1996, Senator Rob Portman cosponsored the Defense of Marriage Act, defining marriage as one man and one woman. In 1999 he voted for a measure prohibiting same-sex couples from adopting children, and in 2011 Portman's ``openly hostile'' record on gay rights led to a mass protest of students at the University of Michigan against his selection as speaker at the graduation ceremony, in response to which his spokesman said that ``Rob believes marriage is a sacred bond between one man and one woman.''  But then in 2013 something happened:  ``I'm announcing today a change of heart [for] gay marriage\dots
My son came to Jane, my wife, and I, told us that he was gay, and that it was not a choice, and that it's just part of who he is, and that's who he'd been that way for as long as he could remember. I've come to the conclusion that \dots [marriage] is something that we should allow people to do \dots and to have the joy and stability of marriage that I've had for over 26 years.'' Portman also mentioned his consultation with former Vice President Dick Cheney, whose own daughter's coming out ``forced him to re-think the issue too, and over time, he changed his view on it.''

This example highlights the potentially important role that familiarity with members of different social groups can have on people's attitudes toward the group and its interests. It also suggests that when thinking about the political salience and influence of a social group, one potentially important and largely understudied factor is the size and characteristics of the \emph{group's} circles of social ties.  In this paper we studied this broader phenomenon, using what we have labeled as groups' social penumbras. With our panel survey we have studied penumbras of 14 groups of varying characteristics and varying political relevance.

We have examined four aspects of these penumbras:  their size, their composition, political attitudes of their members, and the effect on attitudes of entering these penumbras.  These first two attributes (size and composition) can be thought of as aspects of the penumbra and thus, by extension, of the group in question, while the last two attributes characterize the individuals who make up the penumbra. That is, for any group, whether it be gay people, unemployed, Muslims or welfare recipients, we can ask how large is its penumbra and what sorts of people are in that penumbra. We can also ask about the political attitudes of the penumbra's members.

We find that penumbras vary in size, sometimes being much larger than the core group and in other cases less so.  The penumbras of most of the groups we have studied are close to uniformly distributed across the 50 states and are in most cases richer than the average American, which is consistent with the literature finding a positive correlation of income and social network size \citep{zheng2006many}.  The acquaintance penumbras, but not the close-family penumbras, of many of these groups have predictable political slants.

Being in a group's penumbra tends to be positively correlated with positive attitudes on related political issues, but these correlations are high only for some of the groups we have studied.  Using our panel study, we estimate that these correlations represent, at least in part, a causal relationship: Overall, we estimate that entering a group's penumbra has a small but positive effect on attitudes on related political questions.  It is possible that the change in attitudes occurs over a longer period of time than we have examined in our panel, in which case our analysis underestimates the true effect of entering a penumbra.

It is also possible that the largest effect of entering a penumbra is to increase the salience of certain issues rather than to directly change attitudes. In this case, groups' resonance in society may increase in times where its penumbra grows, without public opinion shifting in favor of group-related policies. For example, the fact that Muslims have become a widely discussed social group in many Western countries may mean that more people become aware when their acquaintances are Muslim, in which case its penumbra would grow. Even so, it is not necessarily the case that this larger penumbra would also mean a more pro-Muslim stance in society as a whole: either because members of the penumbra will not become more pro-Muslim, or because this shift would be offset by shifts among those outside the Muslim penumbra. Similarly, there is debate within the pro-choice community as to the potential political effects of more women revealing their abortion histories to their larger social networks.  Further research is needed on the conditions under which changes in the penumbra are likely to translate into more favorable outcomes and political influence for a group.

This study calls for considering penumbras when analyzing the social and politically-relevant characteristics of groups in society. By providing the first measurements of key characteristics of a wide range of penumbras, and demonstrating their potential relevance for explaining policy attitudes---and in some cases, attitude change ---we hope to have advanced this objective. Indeed, we conjecture that penumbras can help account for a wide array of other phenomena, ranging from intergroup relations to variation in media attention to certain phenomena, from fundraising success to the formation of political coalitions. The present study, we hope, lays the foundation for future work on these important issues.

\bibliographystyle{apsr2}

\bibliography{bibpenumbra2}

\pagebreak

\appendix

\section{Survey Items}\label{surveyitems}

\setcounter{table}{0} \renewcommand{\thetable}{A.\arabic{table}}
\begin{itemize}
	
\item  Congress is considering whether to extend the federal unemployment benefits for workers who have exhausted their state unemployment benefits but still cannot find a job. Others worry that such an extension of benefits would add to the national debt. Do you favor or oppose continuing federal unemployment benefits?

\item  Would you support or oppose a law requiring a nationwide ban on the sale of assault weapons?

\item  Do you agree or disagree with the following statement: ``It is the responsibility of the federal government to make sure that all Americans have healthcare coverage''?

\item  From what you have read or heard, is the U.S. spending too much, too little, or about right on public assistance programs to the poor?

\item  Do you think the number of immigrants from foreign countries who are permitted to come to the United States to live should be increased, decreased, or left the same as it is now?

\item  Do you support or oppose allowing illegal immigrants to remain in the country and eventually qualify for U.S. citizenship, if they meet certain requirements like paying back taxes, learning English, and passing a background check?

\item  As a means of preventing terrorist attacks in the United States, would you support or oppose requiring Muslims, including those who are U.S. citizens, to undergo special, more intensive security checks before boarding airplanes in the U.S.?

\item  Do you agree or disagree: Homosexual couples should have the right to marry one another.

\item  Do you think abortion should be legal in all cases, legal in most cases, illegal in most cases, or illegal in all cases?

\item  Which comes closest to your point of view: As a general rule, do you think the United States should be willing to use military force around the world, or the United States should be very reluctant to use military force?

\item  As you may know, the rate of Americans losing their homes through bank foreclosures has risen sharply during the financial crisis. Do you think it would be better for the economy if:
\begin{itemize}
\item A.	The federal government introduces new regulations to prevent this from happening \smallskip
\item  B.	The federal government does not introduce new regulations and instead allows problems in the housing market to be resolved on their own 

\end{itemize} 

\item  Some are calling for the government to provide tax breaks for family expenditures on care provision for elderly family members. Others worry that such tax breaks would add to the national debt. Do you support or oppose providing tax breaks for care provision to the elderly?

\end{itemize}

\section{Group sizes in population}\label{groupsizes}

Figure \ref{fig:bullseye} includes rough estimates of the percentages of adult Americans belonging to each group whose penumbra was asked about in the survey.  The estimates were based on a base of 250 million adults and the following numbers for each group around the time of the first wave of the survey:
\begin{itemize}
\item 11.8 million currently unemployed, from the Bureau of Labor Statistics Economic News Release, 5 Jul 2013:  ``The number of unemployed persons, at 11.8 million, and the unemployment rate, at 7.6 percent, were unchanged in June. Both measures have shown 
little change since February.''
\url{https://www.bls.gov/news.release/archives/empsit_07052013.htm}
\item 10.5 million lost jobs in previous year, from the  Bureau of Labor Statistics Job Openings and Labor Turnover Archived News Releases.
\url{https://www.bls.gov/bls/news-release/jolts.htm#2012}

Source found from Molly's Middle America, 24 Jan 2013:  ``2012 Gross Number of Total Separations:                                                                                49,676,000;
Gross Number of  Layoffs and discharges:                                                                             20,546,000;
Gross Number of Quits:          25,132,000;
Gross Number of `Other' separations:                                                                                   3,997,000.  \url{http://mollysmiddleamerica.blogspot.com/2013/01/how-many-people-lost-their-jobs-in-2012.html}

Also see Louis Jacobson and Molly Moorhead, Polifact, 16 Jan 2012:  ``there is no useful statistic for `Americans (who) have lost their jobs' during a given time period. The labor force is fluid, so people who lose their jobs often move quickly into another one. Instead, economists use the concept of net jobs gained or lost.''  \url{http://www.politifact.com/truth-o-meter/statements/2012/jan/16/mitt-romney/mitt-romney-tweets-more-americans-have-lost-their-/}

\item 5 million members of the National Rifle Association, from Gregory Korte, USA Today, 4 May 2013:  ``Efforts to pass gun-control legislation have only made the National Rifle Association stronger, as the membership rolls now surpass a record 5 million.'' \url{http://www.usatoday.com/story/news/politics/2013/05/04/nra-meeting-lapierre-membership/2135063/}

\item 24\% of Americans own guns, from Drew Desilver, Pew Research Center, 4 June 2013:  ``A Pew Research Center survey conducted in February found that 37\% of households had an adult who owned a gun---24\% said they owned a gun, and 13\% said someone else in their household did.''
\url{http://www.pewresearch.org/fact-tank/2013/06/04/a-minority-of-americans-own-guns-but-just-how-many-is-unclear/}

\item 25\% of Americans with a serious health problem, a rough estimate as ``serious health problem'' has no clear definition.  See the Centers for Disease Control and Prevention Chronic Disease Overview:  ``As of 2012, about half of all adults---117 million people---had one or more chronic health conditions. One of four adults had two or more chronic health conditions.''
\url{http://www.cdc.gov/chronicdisease/overview/}

\item 40 million Americans with no health insurance in 2013, from Kaiser Family Foundation Key Facts about the Uninsured Population, 29 Sep 2016:  ``As of the end of 2015, the number of uninsured nonelderly Americans stood at 28.5 million, a decrease of nearly 13 million since 2013.''
\url{http://kff.org/uninsured/fact-sheet/key-facts-about-the-uninsured-population/}

\item 52.2 million Americans on government welfare, from Shelley K. Irving and Tracy A. Loveless, U.S. Census Bureau, May 2015:  ``In 2012, approximately 52.2 million people, or 21.3 percent of the population, participated in one or more major means-tested assistance programs, on average, each month.'' \url{https://www.census.gov/content/dam/Census/library/publications/2015/demo/p70-141.pdf}.

Source found from Rich Exner, Cleveland.com, 28 May 2015, \url{http://www.cleveland.com/datacentral/index.ssf/2015/05/1_in_5_americans_receive_gover.html}.

\item 4.8 millon adults immigrated to the U.S. in past 5 years, from Jie Zong and Jeanne Batalova, Frequently Requested Statistics on Immigrants and Immigration in the United States, Migration Policy Institute, 26 Feb 2015, ``Twenty-nine percent of the 41.3 million foreign born in the United States in 2013 entered between 2000 and 2009.''
\url{https://web.archive.org/web/20150901173805/http://www.migrationpolicy.org:80/article/frequently-requested-statistics-immigrants-and-immigration-united-states}.  

We calculated $4.8\ {\rm million} = (0.29 * 41.3\ {\rm million\ in\ 10\ years}) * \frac{5\ {\rm years}}{10\ {\rm years}}*\frac{250\ {\rm million \ adults}}{310\ {\rm million\ people}}$.

\item 8.5 million Muslim adults in the U.S., from Frequently Asked Questions about Muslims, Frontline, 2013, ``Estimates range that between five to 12 million Muslims live in the United States,''
\url{https://web.archive.org/web/20131102062347/http://www.pbs.org/wgbh/pages/frontline/shows/muslims/etc/faqs.html}.

Our estimate of 8.5 million is in the middle of that range.

\item 9 millon gay and lesbian Americans, from Gary J. Gates, How Many People are Lesbian, Gay, Bisexual and Transgender?, Williams Institute, April, 2011:  ``Drawing on information from four recent national and two state-level population-based surveys, the analyses suggest that there are more than 8 million adults in the US who are lesbian, gay, or bisexual, comprising 3.5\% of the adult population. In total, the study suggests that approximately 9 million Americans---roughly the population of New Jersey---identify as LGBT. \url{http://williamsinstitute.law.ucla.edu/research/census-lgbt-demographics-studies/how-many-people-are-lesbian-gay-bisexual-and-transgender/}.

\item 5 million Americans who have abortions in the past 5 years, from Induced Abortion in the United States, Guttmacher Institute, ``Approximately 926,200 abortions were performed in 2014, down 12\% from 1.06 million in 2011,'' \url{https://www.guttmacher.org/fact-sheet/induced-abortion-united-states}.  We computed 5 million as approximately 1 million different women per year.

\item 1.4 members of the active military, from United States Armed Forces, Wikipedia, 4 May 2017:  1,429,995 in the United States Army, Marine Corps, Navy, Air Force, Coast Guard, \url{https://en.wikipedia.org/wiki/United_States_Armed_Forces#Personnel}

\item 16.5 million Americans with mortgage under water, from Kathy Orton, 
Number of underwater homeowners continues to decline, Washington Post, 5 Sep 2013:  ``According to the housing data company RealtyTrac, there were 10.7 million U.S. homeowners who owed at least 25 percent more on their mortgages than their homes were worth as of the beginning of September. However, that number has been dropping. It was down from 11.3 million in May and 12.5 million in September 2012.''
\url{http://www.washingtonpost.com/blogs/where-we-live/wp/2013/09/05/number-of-underwater-homeowners-continues-to-decline/}.  We calculated $16.5\ {\rm million} = 11\ {\rm million\ households} * 1.5\ {\rm adults\ per\ household}$.

\item 43.5 million Americans caring for elderly, from Selected Caregiver Statistics, Family Caregiver Alliance, 31 Dec 2012 :
``43.5 million of adult family caregivers care for someone 50+ years of age,''
\url{https://web.archive.org/web/20140511040207/https://caregiver.org/selected-caregiver-statistics}

\end{itemize}

\section{Demographic compositions of penumbras}

\begin{figure}
\centerline{\includegraphics[width=\textwidth]{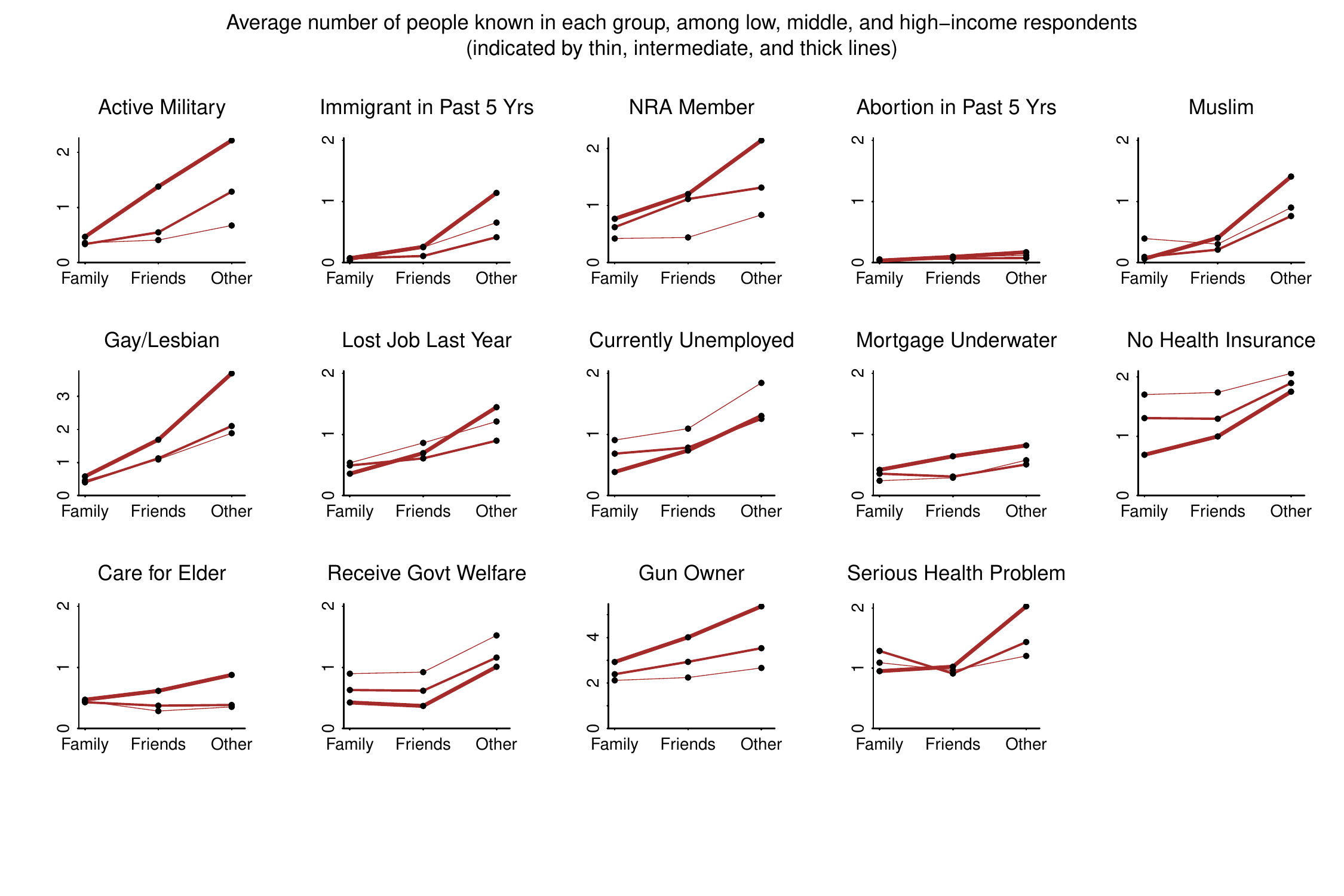}}
\vspace{-.5in}
\caption{\em Average number known in each group among respondents classified by self-reported income.  Wealthier people generally have more acquaintances, with the only exceptions being for poverty-related groups such as welfare recipients or people lacking health insurance.}
\label{fig:income}
\end{figure}

Figure \ref{fig:income} displays the relationship between respondents' income and penumbra membership. As the figure indicates, the unemployed, people with no health insurance, and those receiving welfare are groups whose penumbras are significantly more concentrated among lower-income Americans.

\begin{figure}
\centerline{\includegraphics[width=\textwidth]{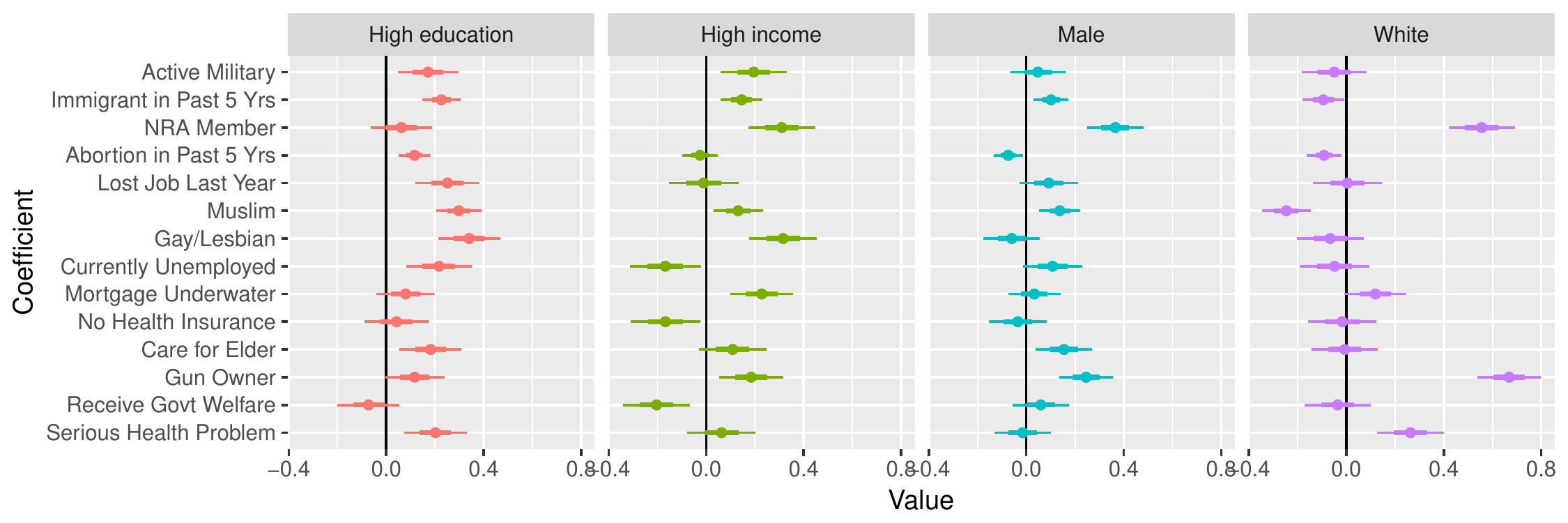}}
\caption{\em Estimated coefficients $\pm$ 2 standard errors, of regressions of penumbra score on indicators for education, income, sex, and race.}
\label{fig:appx3}
\end{figure}

Figure \ref{fig:appx3} shows coefficient estimates, predicting penumbra membership from indicators for education, income, sex, and race.

\subsection{Regressions performed in robustness checks}

\begin{figure}
\centerline{\includegraphics[width=.6\textwidth]{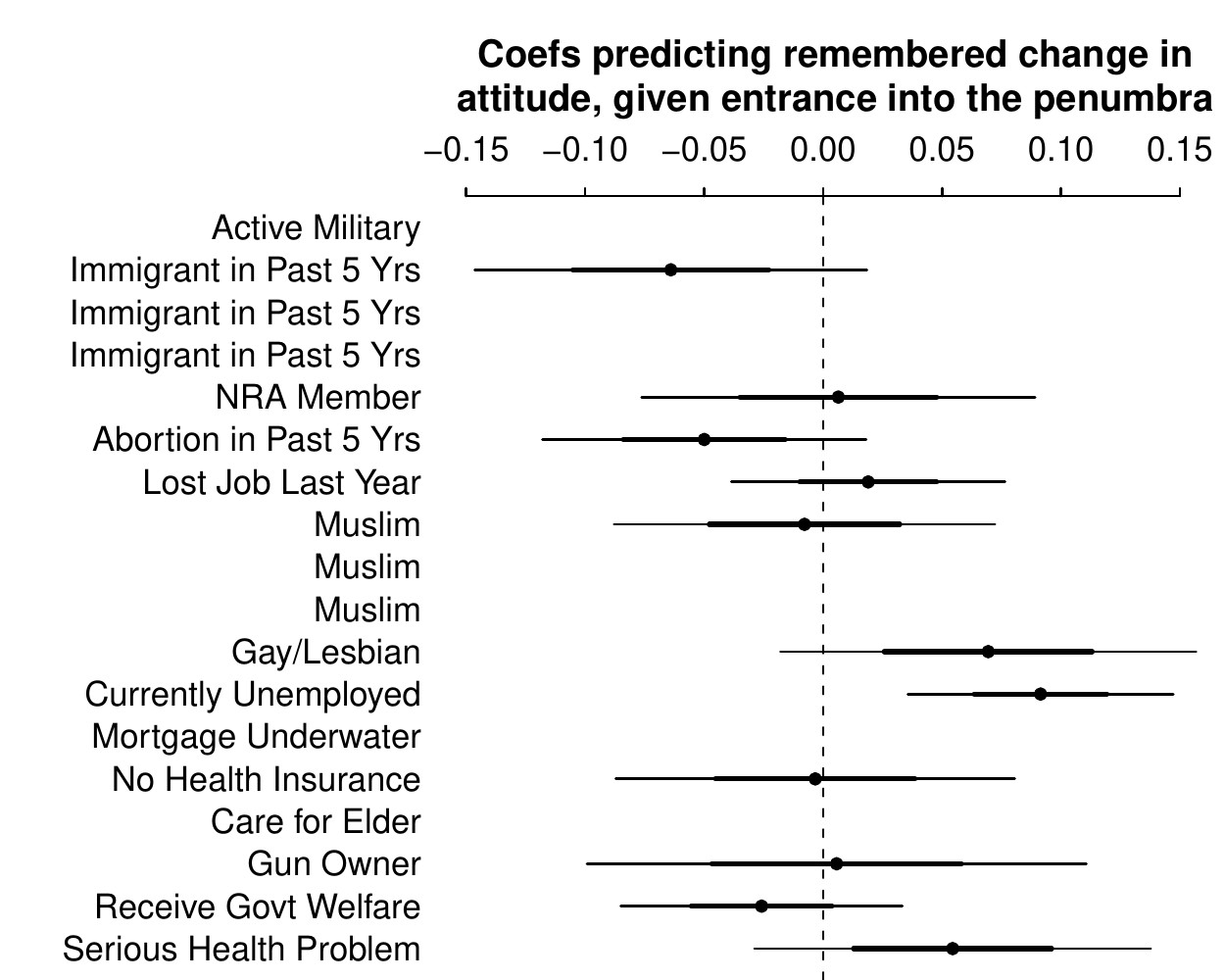}}
\caption{\em Estimated effects of penumbra entry on perceived change in policy views.  The null effects are consistent with the idea that the effects on actual changes shown in Figure \ref{fig:coefs} arose from entering the penumbra rather than simply reflecting concerns about social desirability.}
\label{fig:appx.perceived}
\end{figure}

Figure \ref{fig:appx.perceived} shows the results of the regressions we performed, for each penumbra predicting {\em remembered} changes in policy attitude given entrance into the penumbra.  Compared to the results on actual changes, shown in Figure \ref{fig:coefs}, we see no strong patterns.  The only coefficient that stands out is that for the unemployment penumbra.  The overall lack of pattern in Figure \ref{fig:appx.perceived} serves as a robustness check:  the contrast to Figure \ref{fig:coefs} suggests that the overall positive finding in that earlier figure cannot simply be explained by a demand effect.

Table \ref{figure:cross.regs} shows the coefficients estimated from separate regressions predicting change in policy views given penumbra entry, also controlling for demographics.
Of these $14\times 12=168$ regressions, 18 correspond to issue attitudes and group membership that we predicted ahead of time and 150 are cross-correlations for which we had no expectation of finding effects.  The 150 cross-correlations show no apparent patterns beyond what might be expected by noise, a pattern we also confirm in a formal check.

\begin{table}
\centerline{\begin{small}
\begin{tabular}{l cccccccccccc}
  & $\!\!$Wars$\!\!$ &  $\!\!$Imm+$\!\!$ & $\!\!$Illeg$\!\!$ & $\!\!$Muslm$\!\!$ & $\!\!$Guns$\!\!$ & $\!\!$Abort$\!\!$ &  $\!\!$GayM$\!\!$ & $\!\!$Unemp$\!\!$ & $\!\!$Mortg$\!\!$ &  $\!\!$Hlth$\!\!$ &  $\!\!$Care$\!\!$ &  $\!\!$Welf$\!\!$ \\\hline
Military$\!\!$      &{\bf $-$.04} &$-$.04  &.03 &$-$.03  &.03 &$-$.06  &.01  &.03 &$-$.01 &$-$.02  &.01 &$-$.06  \\
Immigrant $\!\!$ &$-$.01  &{\bf .00}  &{\bf .03}  &{\bf .04} &$-$.01 &$-$.01  &.01  &.07 &$-$.05  &.06  &.11  &.00  \\
NRA$\!\!$     &.01  &.01 &$-$.08 &$-$.04  &{\bf .14} &$-$.04 &$-$.03  &.04 &$-$.09 &$-$.14  &.10 &$-$.10 \\
Abortion $\!\!$   &.04  &.10  &.04 &$-$.05 &$-$.09  &{\bf .04} &$-$.06 &$-$.01  &.03  &.04  &.08  &.00\\
Muslim$\!\!$                 &.02  &{\bf .12}  &{\bf .00}  &{\bf .04}  &.06  &.02  &.08  &.00  &.05  &.05  &.16  &.04  \\
Gay$\!\!$              &.01  &.12  &.02 &$-$.02  &.00  &.07  &{\bf .02} &$-$.01 &$-$.04  &.00  &.04  &.02  \\
Lost Job$\!\!$       &.05  &.07  &.01 &$-$.05 &$-$.06 &$-$.02  &.07  &{\bf .09}  &.04  &.05  &.13  &.06 \\
Unemployed$\!\!$    &.01  &.05 &$-$.04  &.05 &$-$.02  &.01  &.02  &{\bf .11}  &.05  &.01  &.03  &.04  \\
Mortgage $\!\!$    &$-$.01  &.04 &$-$.02 &$-$.03  &.00 &$-$.02  &.07  &.00 &{\bf $-$.04} &$-$.02  &.02  &.06   \\
No Insurance$\!\!$     &.02  &.04 &$-$.08  &.05  &.01 &$-$.04  &.10  &.02  &.00 &{\bf $-$.04}  &.06  &.04   \\
Care Elder$\!\!$          &.06  &.08 &$-$.05  &.00  &.08  &.03 &$-$.02  &.02 &$-$.01 &$-$.04  &{\bf .12}  &.04  \\
Welfare $\!\!$    &$-$.01 &$-$.03  &.02 &$-$.03 &$-$.07 &$-$.03  &.04 &$-$.04  &.00  &.02  &.08  &{\bf .04}  \\  
Gun Owner$\!\!$    &.05 &$-$.06 &$-$.07 &$-$.14  &{\bf .00 } &.04  &.03  &.03  &.02  &.00  &.08  &.01  \\
Health$\!\!$   &.06  &.01  &.06  &.03  &.05  &.04  &.01 &$-$.03 &$-$.06 &{\bf $-$.10} &$-$.02 &$-$.16  
\end{tabular}
\end{small}}
 \caption{\em Coefficients of regressions of change in issue attitudes, given entrance into different penumbras.  The numbers in bold font correspond to the estimates in Figure \ref{fig:coefs}.  The other elements of this matrix, not in bold, are not statistically different from noise.}
\label{figure:cross.regs}
\end{table}

\end{document}